\documentclass[aip, jcp, amsmath, amssymb, reprint]{revtex4-2}
\usepackage{graphicx}
\usepackage{dcolumn}
\usepackage{bm}

\usepackage[utf8]{inputenc}
\usepackage[T1]{fontenc}
\usepackage{mathptmx}
\usepackage{etoolbox}
\usepackage{float}

\usepackage{etoolbox}
\usepackage{hyperref} 

\makeatletter
\def\@email#1#2{%
  \endgroup
  \patchcmd{\titleblock@produce}
    {\frontmatter@RRAPformat}
    {\frontmatter@RRAPformat{%
      \produce@RRAP{#1\href{mailto:#2}{#2}}%
    }\frontmatter@RRAPformat}
    {}
    {\PackageError{correspondence}
      {Could not insert corresponding-author email}
      {Check the REVTeX title-block customizations.}}%
}
\makeatother

\usepackage{xcolor}

\begin{document}

\preprint{AIP/123-QED}

\title[]{Numerical integration of intracule pair densities with optimized multicenter grids}
\author{M. Ylla}
\affiliation{Donostia International Physics Center (DIPC), 20018 Donostia, Euskadi, Spain}
\affiliation{Kimika Fakultatea, Euskal Herriko Unibertsitatea (EHU), P.K. 1072, 20080 Donostia, Euskadi, Spain}
\author{J. M. Ugalde}
\affiliation{Donostia International Physics Center (DIPC), 20018 Donostia, Euskadi, Spain}
\affiliation{Kimika Fakultatea, Euskal Herriko Unibertsitatea (EHU), P.K. 1072, 20080 Donostia, Euskadi, Spain}

\author{E. Matito\textsuperscript{*}}
\email[\textsuperscript{*}Corresponding author: ]{ematito@dipc.org}
\affiliation{Donostia International Physics Center (DIPC), 20018 Donostia, Euskadi, Spain}
\affiliation{Kimika Fakultatea, Euskal Herriko Unibertsitatea (EHU), P.K. 1072, 20080 Donostia, Euskadi, Spain}
\affiliation{Ikerbasque Foundation for Science, Plaza Euskadi 5, 48009 Bilbao, Euskadi, Spain}

\author{E. Ramos-Cordoba\textsuperscript{\ensuremath{\dagger}}}
\email[\textsuperscript{\ensuremath{\dagger}}Corresponding author: ]{eloy.ramos@iqac.csic.es}
\affiliation{Institute for Advanced Chemistry of Catalonia (IQAC), CSIC, Jordi Girona 18-26, 08034 Barcelona, Spain}

\date{\today}

\begin{abstract}
 The analysis of pair-density intracule functions offers deep insight into electron correlation effects, yet the routine calculation of these quantities for extended systems has been hindered by the prohibitive cost of numerical integration or by the limitations of analytical methods. In contrast to the latter, which are restricted to Gaussian basis functions, the present approach is applicable to arbitrary basis functions. We present an efficient and scalable multicenter integration scheme for the calculation of radial intracule densities and their associated moments. By adapting the topological fuzzy Voronoi cell (TFVC) formalism, we define a robust, automated partitioning of the intracule space. Benchmarks on the linear alkane series reveal that standard single-center grids suffer from severe convergence issues as molecular size increases, failing to capture intracule density peaks located at internuclear separations. In contrast, the proposed multicenter approach restores high precision (relative errors in $V_{ee}$ below 0.02\%) with a fraction of the points required by single-center methods. Topological analysis confirms that the multicenter scheme naturally clusters quadrature points around the secondary maxima of the intracule density, effectively resolving the \textit{blind spots} inherent to single-center grids. Beyond computational efficiency, this spatial partitioning facilitates the decomposition of global two-electron properties into chemically meaningful short-range (intra-atomic) and long-range (inter-atomic) contributions. Furthermore, by coupling the multicenter grid with a kernel density estimation technique, we reconstruct the continuous radial intracule probability distribution, $I(u)$, at large interelectronic distances, completely bypassing the need for the massive angular grids required by standard surface integration. Validation against the S22 dataset confirms the reliability of the method across diverse covalent and non-covalent interactions, establishing it as a general tool for high-precision intracule analysis in polyatomic systems. 
\end{abstract}

\maketitle

%


\section{Introduction}

Electron correlation, defined as the collective interaction between electrons beyond the mean field approximation,\cite{lowdin:55pr} remains as one of the central challenges in quantum chemistry. Accurately describing these interactions is indispensable for the correct prediction of chemical phenomena, including reaction barriers, radical stability, and non-covalent interactions.\cite{hattig:12cr} Consequently, the development of electronic structure methods for the efficient capture of correlation energy is a primary objective for theoretical chemists. To design functional and wavefunction approximations, one must first fundamentally understand how electron correlation manifests itself in the spatial distribution of electrons and how it affects global properties such as the total energy. To this end, our group has devoted considerable effort to developing measures of dynamic and nondynamic correlation,\cite{valderrama:01jpb,mercero:03bch,ramos-cordoba:16pccp,ramos-cordoba:17jctc} which can subsequently be used to design new electronic-structure methods\cite{paulau:25jctc} and multireference diagnostics.\cite{xu:25jcp,xu:24jctc}

A rigorous and physically intuitive tool for analyzing these effects is the radial intracule density $I(u)$, obtained by reduction of the pair density. This probability density function describes the likelihood of finding two electrons separated by a scalar distance $u$, effectively projecting the multidimensional electron-pair density onto a single coordinate. The recourse to the concept of electron-pair distribution function was pioneered by Coulson and Neilson \cite{Coulson1961} in their seminal analysis of the Helium atom, where they demonstrated a direct link between the mean interelectronic separation and the error in the calculated energy. This framework was later formalized by Coleman \cite{Coleman_1967} within the context of reduced density matrices (RDMs), establishing the intracule as a useful descriptor of electron-pair separation and electron correlation.

The primary utility of the intracule density lies in its ability to visualize electron correlation through the so-called \textit{holes}, which result from the difference of two intracule densities. The Coulomb Hole, obtained by comparing the correlated and Hartree-Fock intracule densities, characterizes the depletion of the probability of finding electron pairs at short interelectronic distances due to correlation. This analysis has been applied to atoms \cite{Smith_1971, Boyd_1973,rodriguez:19co} and to the study of angular correlation effects.~\cite{boyd1982angular} Furthermore, Boyd and Coulson extended this definition to parallel-spin electrons to characterize the Fermi hole,\cite{Boyd_1974} providing a direct spatial visualization of the Pauli exclusion principle.

Beyond atomic systems, intracule analysis has proven vital for understanding bonding\cite{rodriguez:18jpca} and correlation in polyatomic molecules. Boyd, Sarasola, and Ugalde~\cite{Boyd_1988, Ugalde_1985, Sarasola_1990} utilized intracules to map the topological features of the Fermi and Coulomb holes, explicitly highlighting the inadequacy of restricted HF wavefunctions in describing molecular dissociation. Similarly, Dominguez et al.~\cite{Dominguez_1992} explored the basis-set dependence of pair densities in $\text{CO}_2$. More recently, the topological features of intracules have been used to disentangle dynamic from nondynamic correlation effects, \cite{Valderrama:02bookchapter, valderrama:01jpb, via-nadal:19jpcl, via-nadal:22jcp} and to develop entirely new theoretical frameworks for correlation energy estimation. A prominent example is Intracule Functional Theory (IFT) and the associated $\Omega$-intracule models proposed by Gill and co-workers, \cite{Gill_2003, Gill_2006, Bernard_2008} which construct correlation functionals directly from the intracule of the pair density rather than the one-electron density. 

Despite their theoretical importance, the calculation of radial intracules for large molecular systems has historically been hindered by their high computational cost. The calculation first requires obtaining the intracule density by reducing the pair density, which, for correlated wavefunctions, is already costly. The radial intracule is then obtained from the intracule density by integration over the angular coordinates. Finally, an additional integration is required in order to compute moments and other global two-electron properties, which further increases the computational cost. Different computational strategies have been optimized depending on the target quantity: the intracule, the radial intracule, or global two-electron properties.

Two main strategies have emerged to address these challenges. The first is the analytical evaluation of intracule integrals. This approach was initially pioneered by Ugalde, Sarasola, and Boyd,~\cite{Sarasola_1990, Ugalde_1991_Algorithm} who developed the foundational analytical formulations for polyatomic molecules, and was later highly optimized by Gill and co-workers through the adaptation of the PRISM algorithm.~\cite{ONeill_Gill_2003} Although analytically exact and computationally efficient, this strategy involves an analytical integration over the solid angle that averages out the angular information. Consequently, while these analytical methods permit the integration of radial intracules up to an arbitrary scalar distance $u$, they cannot decompose properties along interatomic vectors. In contrast, our multicerenter spatial partitioning preserves this multidimensinal information, enabling the decomposition of global two-electron properties into localized, vector-separated contributions.

The second approach relies on a tailor-made numerical integration scheme. A primary theoretical advantage of numerical integration is its universality: the multicenter grid itself can be applied to any wavefunction form or basis set. To carry out such an integration, however, one must first evaluate the intracule density at multiple grid points. For this purpose, we employ the algorithm proposed by Cioslowski and Liu to compute the intracule density at a given point in space, using two-electron integral screening to reduce the computational cost. However, the adequacy of the numerical integration depends strongly on the quality of the integration grid. Owing to the topology of the intracule density, an origin-centered radial and angular quadrature is not sufficient to resolve its sharp features, including the cusp at zero and the features occurring at interatomic distances. \cite{via-nadal:19jpcl} Very large angular quadratures are therefore required to obtain reliable results, rendering the calculation computationally prohibitive for large molecules.

Because this specific point-evaluation algorithm is explicitly formulated for Gaussian basis functions, our current implementation remains restricted to them. Extending this numerical approach to evaluate non-Gaussian basis sets would require the development of alternative algorithms to feed the intracule values into the grid.

To efficiently improve the numerical integration of the intracule, the grid distribution must be optimized. In the context of density functional theory (DFT), this problem was addressed by Becke’s multicenter integration scheme.\cite{becke:88jcp} Becke recognized that the electron density is dominated by cusps at the atomic nuclei and proposed partitioning the integrand into single-center atomic contributions. Although this approach may fail for properties that are not sufficiently atom-centered,\cite{laqua:18jcp} such as some high-order energy derivatives,\cite{zalesny:19jctc,sitkiewicz:22jpcl,sitkiewicz:23pccp,sitkiewicz:24jctc} it enables the use of locally optimized grids that efficiently sample regions of high density, achieving high accuracy with only a fraction of the grid points required by standard single-center quadratures.

Here, we propose a novel multicenter integration scheme specifically adapted to the unique topology of intracule densities. Building directly on Becke’s spatial partitioning framework,\cite{becke:88jcp} we redefine the integration centers as interatomic displacement vectors and use the topological \textit{fuzzy} Voronoi cell (TFVC) formalism\cite{salvador:13jcp} to construct the corresponding integration weights. This adaptation preserves the universal applicability of numerical integration while achieving substantially improved computational efficiency.

The remainder of this paper is organized as follows. In Section II, we briefly review the theoretical framework of intracule functions and outline Cioslowski and Liu’s screening formalism for accelerating their numerical evaluation. We then describe our adaptation of Becke’s multicenter integration scheme, based on the TFVC formalism, to optimally distribute radial and angular quadrature points across intracule space. Finally, in Section IV, we present benchmark results for linear alkanes and the diverse S22 dataset, validating the accuracy, scalability, and chemical utility of the proposed grid methodology. We also introduce 
the spatial decomposition of two-electron properties and the reconstruction of long-range radial distributions through Gaussian smearing.

\section{Theory}

\subsection{The Intracule Density and its Moments}

The exact two-electron density matrix (2-RDM) contains the full quantum mechanical description of the average electron-pair in a molecular system. More specifically, its diagonal ---the so-called pair density, $\rho_2(\mathbf{r}_1,\mathbf{r}_2)$--- represents the spatial probability of simultaneously finding one electron at position $\mathbf{r}_1$ and another at $\mathbf{r}_2$, regardless the position of the $N-2$ remaning electrons. To study electron correlation, it is convenient to separate these six coordinates into a center-of-mass (extracule) coordinate, $\mathbf{R}=(\mathbf{r}_1+\mathbf{r}_2)/2$, and a relative (intracule) coordinate, $\mathbf{u}=\mathbf{r}_1-\mathbf{r}_2$. The intracule density, $I(\mathbf{u})$, is obtained by integrating the pair density over all extracule coordinates $\mathbf{R}$, 
\begin{equation}
I(\mathbf{u})=\int d\mathbf{R} \,\,\,\rho_2(\mathbf{R}+\frac{\mathbf{u}}{2},\mathbf{R}-\frac{\mathbf{u}}{2}),    
\end{equation}
and it represents the 3D probability density of finding any two electrons separated by the vector $\mathbf{u}$. 

While $I(\mathbf{u})$ contains the full angular dependence of the interelectronic vectors, it is often convenient to reduce it to the radial intracule density, $I(u)$, via an angular integration,
\begin{equation}\label{eq:angular}
    I(u) = u^2\int d\Omega_{\mathbf{u}} \,\,\, I(\mathbf{u}),
\end{equation}
where $d\Omega_u=sin(\theta)d\theta d\phi$ is the solid angle of the vector $\mathbf{u}$. This function describes the probability of finding two electrons at a scalar distance $u$. Depending solely on the interelectronic distance, it is easier to visualize and reveals the radial distribution of the electron-pair separation for a given wavefunction.

One important utility of the radial intracule density lies in its moments, which yield fundamental global properties of the system through a an additional radial integration. The moment of order $n$ is defined as
\begin{equation}\label{eq:moments}
    \langle u^n \rangle = \int_0^{\infty} du \,\,\, u^{n} I(u).
\end{equation}
The $n=0$ moment recovers the number of electron pairs of the molecule, given by $N(N-1)/2$, where $N$ denotes the number of electrons and the $n=-1$ moment yields the electron-electron repulsion energy, $V_{ee} = \langle u^{-1} \rangle$. Higher-order moments provide information about the compactness and diffuseness of the electron-pair cloud. Furthermore, evaluating Eq.~\ref{eq:moments} over a restricted domain yields range-separated energetic contributions, which are useful, for example, for describing long-range non-covalent interactions.\cite{via-nadal:17pra,via-nadal:19jpcl} Therefore, the accurate and efficient numerical integration of intracule densities is essential for analyzing electron correlation \cite{rodriguez:18jpca, Boyd_1988, Ugalde_1985, Sarasola_1990, Dominguez_1992, Valderrama_1999, valderrama:01jpb, via-nadal:19jpcl, via-nadal:22jcp} and informingly developing new electronic-structure methods. \cite{paulau:25jctc,Gill_2003, Gill_2006, Bernard_2008}

\subsection{Numerical Evaluation of Intracule Integrals}

Fully analytical methods can evaluate this entire sequence ---from the six-dimensional pair density directly to the final intracule moments--- bypassing the need for explicit angular or radial grids.~\cite{ONeill_Gill_2003} While analytically exact, this strategy is generally restricted to Gaussian basis sets and specific operators. To maintain universality and flexibility for arbitrary correlated wavefunctions, we instead employ a hybrid analytical-numerical approach.

The first step is the evaluation of the intracule density, $I(\mathbf{u})$, at a given interelectronic vector. This evaluation requires integration over the extracule coordinates, which is performed analytically. Here, we employ the efficient algorithm developed by Cioslowski and Liu.~\cite{cioslowski1996fast} In this algorithm, the intracule is evaluated as a contraction of the primitive representation of the two-electron density matrix, $\Gamma_{ijkl}$, with primitive basis-function integrals, $[ij|kl]_\mathbf{u}$, 
\begin{equation}\label{eq:cioslowski}
    I(\mathbf{u}) = \sum_{ijkl} \Gamma_{ijkl} [ij|kl]_\mathbf{u},
\end{equation}
where, 
\begin{equation}\label{eq:ciosintegral}
   [ij|kl]_\mathbf{u}= \int\int d{\mathbf{r_1}}d\mathbf{r_2}\phi^*_i(\mathbf{r_1})\phi^*_j(\mathbf{r_2})\phi_k(\mathbf{r_1})\phi_l(\mathbf{r_2})\delta(\mathbf{r_2}-\mathbf{r_1}-\mathbf{u}),
\end{equation}
and $\phi(\mathbf{r})$ stand for primitive Gaussian functions.

The evaluation of this function scales formally as $O(M^4)$, where $M$ is the total number of primitive functions. However, Cioslowski and Liu’s algorithm employs a two-stage, grid-independent screening protocol to reduce the number of primitive quartets that must be evaluated for a given $\mathbf{u}$ vector. A first screening eliminates negligible primitive quartets based on the magnitude of $\Gamma_{ijkl}$ and primitive basis function overlaps. In the second screening, an upper bound of the grid dependent part is computed, eliminating an extra amount of quartets that do not enter in the final evaluation of the function. In this way, we avoid many evaluations of primitive quartets that do not contribute to the value of the intracule at any point.

With the analytical intracule values available, the second step is the numerical integration over the solid angle to obtain the radial intracule value $I(u_i)$ at a specific distance $u_i$. This is approximated via an angular quadrature: 
\begin{equation}\label{eq:num_angular}
    I(u_i) \approx u^2_i\sum_{j=1}^{N_{ang}} w^\text{ang}(\Omega_j) I(u_i, \Omega_j),
\end{equation}
where $u_i \text{ and } \Omega_j$ are radial and angular coordinates, respectively, $N_{ang}$ is the total number of points on the angular grid, and  $w^\text{ang}(\Omega_j)$ are the weights of the angular nodes.

The final step consists of evaluating the moments of Eq.~\ref{eq:moments} by applying a numerical radial quadrature over the discrete distance grid.
\begin{equation}\label{eq:numercial_moments}
     \langle u^n \rangle \approx \sum_{i=1}^{N_\text{rad}} w^\text{rad}(u_i) I(u_i) u_i^{n},
\end{equation}
where $w^\text{rad}(u_i)$ are the corresponding radial quadrature weights and $N_{rad}$ the total number of radial nodes.
Despite the analytical prescreening introduced by Cioslowski and Liu for Eq.~\ref{eq:cioslowski}, the numerical evaluation of global properties scales as $N_\text{rad} \times N_\text{ang} \times M^4$.

Furthermore, the standard single-center spherical grid, centered at the origin, suffers from a severe geometric limitation: as the interelectronic distance increases, a fixed angular grid becomes progressively too sparse to accurately sample the expanding spherical surface. Suppressing the resulting numerical noise and resolving long-range interactions would require an impractically large $N_\text{ang}$ grid, thereby creating a major computational bottleneck. 

\subsection{Multicenter Partitioning of the Intracule Space}

A multicenter partitioning scheme is better suited to the multimodal nature of the intracule density, which exhibits maxima at vectors corresponding to interatomic separations, $\mathbf{u}_{AB}=\mathbf{r}_A-\mathbf{r}_B$, $\mathbf{r}_A \text{ and }\mathbf{r}_B$ being the position of atoms A and B, respectively. Such a scheme will permit a more efficient distribution of the integration grid points than the single-center spherical expansion.

We adapt the multicenter numerical integration scheme originally proposed by Becke for real-space electron densities.\cite{becke:88jcp} Instead of placing integration centers at atomic coordinates, we define the set of integration centers in the intracule space as the interatomic displacement vectors, plus the origin ($\mathbf{u}=0$) which captures the short-distance electron pairs, mainly localized around each of the nuclei. To ensure numerical stability, the centers separated by a distance smaller than a predefined threshold (0.1 bohr) are merged into a single center. 

The total integration over the intracule space is exactly partitioned into a sum over these centers, 
\begin{equation}
    \int d\mathbf{u} \,\,\, F(\mathbf{u}) = \sum_{C} \int d\mathbf{u} \,\,\, w_C(\mathbf{u}) F(\mathbf{u}) ,
    \label{eq:partitioning}
\end{equation}
where $F(\mathbf{u})$ is the property to be integrated (e.g. $I(\mathbf{u})/u$ for $V_{ee}$) and $w_C(\mathbf{u})$ are smooth weight functions that satisfy $\sum_Cw_C(\mathbf{u)}=1$ for all $\textbf{u}$. 

The weight functions $w_C(\mathbf{u})$ define \textit{fuzzy} Voronoi cells around each center. The weight for center $C$ is constructed by evaluating the cell function $P_C$,
\begin{equation}
w_C(\mathbf{u}) = \frac{P_C(\mathbf{u})}{\sum_D P_D(\mathbf{u})}, \quad P_C(\mathbf{u}) = \prod_{D \neq C} s_{CD}(\mu_{CD}),
\end{equation}
where $\mu_{CD}$ is the hyperbolic coordinate between centers $C$ and $D$, 
\begin{equation}
\mu_{CD} = \frac{||\mathbf{u} - \mathbf{u}_C|| - ||\mathbf{u} - \mathbf{u}_D||}{||\mathbf{u}_C - \mathbf{u}_D||},
\end{equation}
which ranges from $-1$ (at center $C$) to $+1$ (at center $D$). The switching function $s_{CD}(\mu_{CD})$ ensures a smooth transition across the boundary between the two centers. Following Becke, \cite{becke:88jcp} we employ the iterated polynomial step function,
\begin{equation}\label{eq:scd}
s_{CD}(\mu_{CD}) = \frac{1}{2} \left[ 1 - f_k(\mu_{CD}) \right],
\end{equation}
where $f_k(\mu)$ is generated by iterating the function $f(\mu) = \frac{3}{2}\mu - \frac{1}{2}\mu^3$ exactly $k$ times. In this work, we set $k=4$ (i.e., $f_4(\mu) = f(f(f(f(\mu))))$). 

In this formulation, the cell boundary ($s_{CD} = 0.5$) corresponds to $\mu_{CD}=0$, which geometrically defines the perpendicular bisector plane between the centers.
However, the integration centers in the intracule space differ markedly in size, that is, in the number of electron pairs they represent, which depends on the particular pair of atoms involved. Consequently, the standard bisection used in conventional Voronoi partitioning is not appropriate.

Although Becke proposed a polynomial shift to account for heteronuclear systems,\cite{becke:88jcp} this correction is known to fail, or even worse, to become non-monotonic, when the size disparity between centers is very large.
To robustly handle the arbitrary size differences present in the intracule space, we employ the TFVC formalism to define the weights in Eq.\ref{eq:partitioning}.~\cite{salvador:13jcp} The boundary between centers $C$ and $D$ is shifted according to a relative size ratio, $\chi_{CD}$. In our adaptation, this ratio is defined exclusively from the electron-pair population associated with each center,
\begin{equation}\label{eq:chi}
\chi_{CD} = \sqrt[3]{\frac{\mathcal{N}_C}{\mathcal{N}_D}}.
\end{equation}
The population $\mathcal{N}_C$ is defined as the total number of electron pairs assigned to center $C$. Because the interatomic center $C$ in intracule space may be constructed by merging a set of nearly equidistant interatomic pairs, $k=(A,B)$, is given by the sum of the products of the corresponding formal atomic populations: $\mathcal{N}_C = \sum_{k \in C} N_{A(k)} N_{B(k)}$. For the center at the origin ($\mathbf{u}=0$), which captures all intra-atomic pairs, the population is $\sum_A N_A(N_A-1)/2$. The cube root in Eq.~\ref{eq:chi} ensures dimensional consistency by converting the pair population, which is treated as a volumetric analogue, into a linear scaling factor consistent with the spatial coordinates $\mu$.

This ratio dictates the boundary shift via the TFVC transformation:
\begin{equation}
\nu_{CD} = \frac{1-\phi}{1+\phi}, 
\end{equation}
where
\begin{equation}
\phi=\chi_{CD}\left(\frac{1-\mu_{CD}}{1+\mu_{CD}}\right).
\end{equation}

The variable $\nu_{CD}$ replaces $\mu_{CD}$ as the argument of the switching function in Eq. \ref{eq:scd}, $s_{CD}(\nu_{CD})$, thereby systematically displacing the \textit{fuzzy} boundary away from highly populated centers and toward sparsely populated ones. This construction ensures an appropriate sampling volume for each integration center.

With the total integral partitioned into single-center contributions, we employ a mapped Gauss-Legendre quadrature for the radial coordinate. \cite{Press_2007, becke:88jcp} The standard nodes $x_i \in [-1,1]$ are mapped to the semi-infinite radial distance  $u_i \in [0, \infty)$ through the algebraic transformation introduced by Becke,
\begin{equation}
    u_i = \alpha_C \left(\frac{1+x_i}{1-x_i}\right),
\end{equation}
where $\alpha_C$ is a center-specific scaling parameter that controls the radial distance enclosing half of the grid points. 

 To adapt the radial grid to the physical diffuseness of the electron pairs associated with a given center, we derive $\alpha_C$ from the empirical Bragg-Slater atomic radii, $R^0$. Using the constituent atomic pairs $k$ mapped onto center $C$, we compute $\alpha_C$ as their population-weighted average, 
\begin{equation}
\alpha_C = \frac{1}{\mathcal{N}_{C}} \sum_{k } \mathcal{N}_k \left( \frac{R_{A(k)}^0 + R_{B(k)}^0}{2} \right),
\end{equation}
where $\mathcal{N}_k=N_{A(k)}N_{B(k)}$ is the population of the specific pair $k$. Other choices of $\alpha_C$ are possible, but our experience with a selected set of examples indicates that this definition provides good results. 

For the center at the origin, the scaling parameter $\alpha_0$ is computed as the population-weighted average of the radii of all atoms in the molecule, ensuring that the grid adequately samples the origin cusps of the heaviest elements present. This guarantees that the radial grid dynamically adapts to the specific chemical nature of each center without requiring user intervention or empirical tuning.

\subsection{Radial Intracule Reconstruction via Kernel Density Estimation}

As we shall see, the multicenter quadrature scheme described above provides an efficient and robust numerical framework for computing scalar properties. However, the radial intracule $I(u)$ cannot be obtained directly from this procedure. 

In single-center schemes, $I(u)$ is obtained by explicit angular integration over a spherical shell of radius $u$:
\begin{equation}
    I(u) = u^2 \int d\Omega_u \,\,\,I(\mathbf{u})=u^2 \int_0^{2\pi}d\phi \int_0^{\pi}\sin\theta \, d\theta \,\,\,  I(u, \theta, \phi) \, .
\end{equation}
Hence, all grid points are distributed around the only available center, the origin, and the assignment of interelectronic distances, $u$, is therefore straightforward. Such grids are typically constructed from spherical quadratures, for example Gauss–Lebedev angular grids, defined on fixed radial shells. As a result, all angular points on a given shell contribute to the same value of $u$.

In a multicenter framework, except for the trivial single-center case corresponding to an atom, grid points are distributed around centers that are generally displaced from the origin. Consequently, the distance $u=|\mathbf{u}|$ of a grid point to the origin is no longer determined solely by its local radial shell. The grid therefore lacks a simple global fixed-shell structure that can be directly exploited to reconstruct the radial intracule by standard spherical quadrature.

To overcome this limitation and reconstruct the continuous function $I(u)$ from the discrete multicenter data, we adopt a Gaussian smearing technique, namely the kernel density estimation (KDE). This approach is mathematically analogous to the smearing methods routinely employed in solid-state physics for evaluating the Density of States (DOS) from discrete $k$-point sampling of the Brillouin zone. \cite{Methfessel_1989, Martin_2004}

Instead of integrating over a spherical surface, we treat the multicenter grid as a discrete volume sampling of the intracule density. Each grid point $i$ represents a differential volume element associated with a total integration weight, ${w(\mathbf{u_i})}$, which is a product of the TFVC, the radial, and the angular weights. The total probability contained within this volume element, ${w(\mathbf{u_i})} I(\mathbf{u}_i)$, is then projected onto the continuous radial probability distribution, $I(u)$, using a normalized Gaussian kernel $K(u, u_i)$:
\begin{equation}
    I(u) \approx \sum_{i=1}^{N_{grid}}  w(\mathbf{u_i})  I(\mathbf{u}_i) K(u - u_i) ,
\end{equation}
where $u_i = |\mathbf{u}_i|$ is the distance of the $i$-th grid point from the origin, and the kernel is defined as:
\begin{equation}
    K(u - u_i) = \frac{1}{\sigma\sqrt{2\pi}} \exp\left( -\frac{(u - u_i)^2}{2\sigma^2} \right)
\end{equation}

Here, $\sigma$ is a broadening parameter that must be chosen to balance resolution with numerical stability; we fix this value to 0.15. The broadening effectively bridges the gaps between discrete grid points in the radial direction, converting the sparse sampling into a smooth continuous function.

This method is strictly conservative; integrating the reconstructed $I(u)$ over all $u$ recovers the exact total number of electron pairs \textit{computed by the underlying quadrature}.
The primary advantage of this reconstruction is its ability to resolve long-range features. While single-center angular grids become increasingly sparse at large distances (the distance between angular grid points scaling as $1/u^2$), the multicenter grid places high-density clusters of points directly at the interatomic separations $\mathbf{R}_{AB}$. The smearing method naturally aggregates these weights into the correct radial bin, allowing for the precise resolution of long-range correlation effects, such as dispersion peaks and Coulomb holes, which are often missed by standard angular integration schemes.

\section{Computational Details}

The multicenter integration algorithm described in this work has been implemented in the \textsc{inca} code, written in \textsc{fortran} and available in GitHub. \cite{incacode} For the numerical quadratures, we employ the Lebedev-Laikov grids for the angular integration, \cite{Lebedev} and the mapped Gauss-Legendre quadrature for the radial coordinate. \cite{Press_2007, becke:88jcp} To determine the optimal balance between computational cost and accuracy, a wide range of angular and radial grid sizes were benchmarked; the specific grid combinations used are detailed in the respective Results subsections. In all the systems studied, atomic pairs involving a hydrogen atom are not considered as integration centers. 

The HF level of theory and STO-3G basis set were chosen to facilitate extensive benchmarking and efficient handling of the two-electron density matrices. It should be noted that the primary objective of this work is to validate the numerical precision of the integration algorithm (specifically, its ability to recover the exact number of electron pairs and reproduce the electron-electron potential) rather than to compute highly accurate correlation energies. 

All the calculations were performed with Gaussian16,~\cite{g16} and the resulting formatted checkpoint files were processed using the \textsc{DMN} code (an in-house program developed by Matito and co-workers)\cite{matito:07fd,feixas:14jctc} to generate the second-order reduced density matrix in primitive basis functions and the basis set information. These files served as the input for \textsc{inca} to perform the intracule integrations.

Finally, it is worth noting that, in section C, we used Gaussian16\cite{g16} and Knowles's FCI program,\cite{knowles:89cpc} to compare FCI to HF calculations with aug-cc-pVTZ basis sets, in order to observe the dispersion interactions that are related with long-range dynamic electron correlation. 

\section{RESULTS AND DISCUSSION}

\subsection{Method performance}
\begin{figure}[h]
    \centering
    \includegraphics[width=1\linewidth]{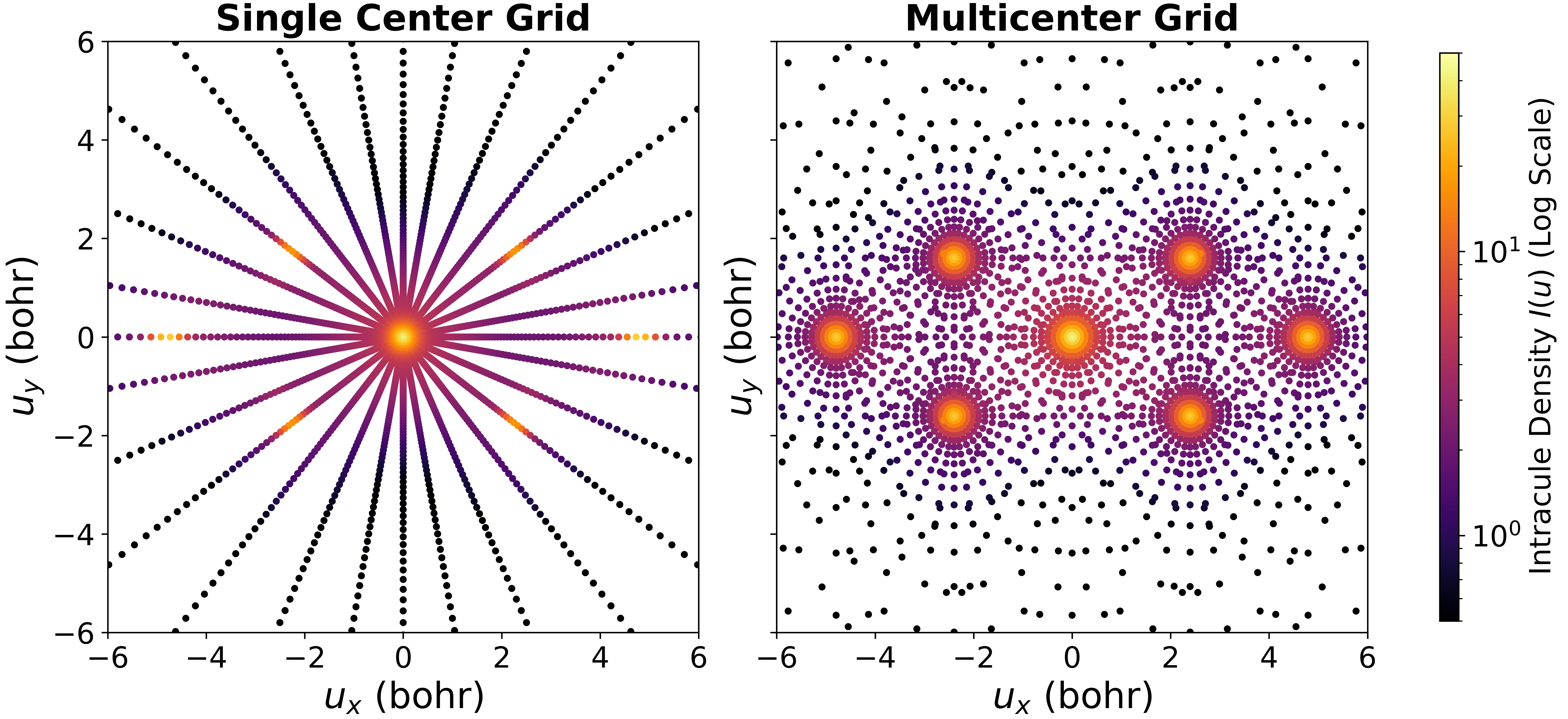}
    \caption{Cross-sectional visualization of the numerical integration grids for the propane molecule. The quadrature points are projected onto the $xy$-plane and colored by the logarithmic magnitude of the intracule density. Left: Single-center grid ($N_{rad}=210, N_{ang}=590$). Right: Multicenter TFVC grid ($N_{rad}=30, N_{ang}=590$).}
    \label{fig:grid_comparison}
\end{figure}

To illustrate the fundamental structural advantage of the multicenter expansion over the single-center approach, it is instructive to visualize the spatial distribution of the quadrature points relative to the intracule density. Figure \ref{fig:grid_comparison} presents a cross-sectional projection of the integration grids in the $xy$-plane for the propane molecule, with each point colored by the logarithmic magnitude of the intracule density. In both cases, a comparable total number of grid points is used ($N_{rad}=210$, $N_{ang}=590$ for the single center, and $N_{rad}=30$, $N_{ang}=590$ per center for the multicenter).

The left panel displays the single-center grid. Despite employing a dense radial quadrature, the distribution of points exhibits a characteristic sparse sampling of the bond regions. While the central intra-atomic pair peak (at $u_x=0, u_y=0, u_z=0$) is sampled with high fidelity, the grid density decays rapidly with distance. The grid samples these high-density regions with very few points, which fundamentally limits convergence and causes the severe $V_{ee}$ integration errors that will be demonstrated in the subsequent benchmarking sections.

In contrast, the right panel demonstrates that the multicenter TFVC grid features a topology that naturally mimics the electron pair distribution. By defining integration centers at internuclear vector positions, the scheme ensures an effective clustering of points at the interpair density peaks.

We assess the performance of the adapted multicenter integration scheme by comparing its accuracy and efficiency with those of a standard single-center quadrature. The primary validation metric is the electron–electron repulsion energy, $V_{ee}$ obtained from the radial intracule and compared with the reference value computed from two-electron integrals using the standard Obara–Saika recurrence scheme.~\cite{obara:86jcp}

\begin{figure}[h]
\centering
\includegraphics[width=1\linewidth]{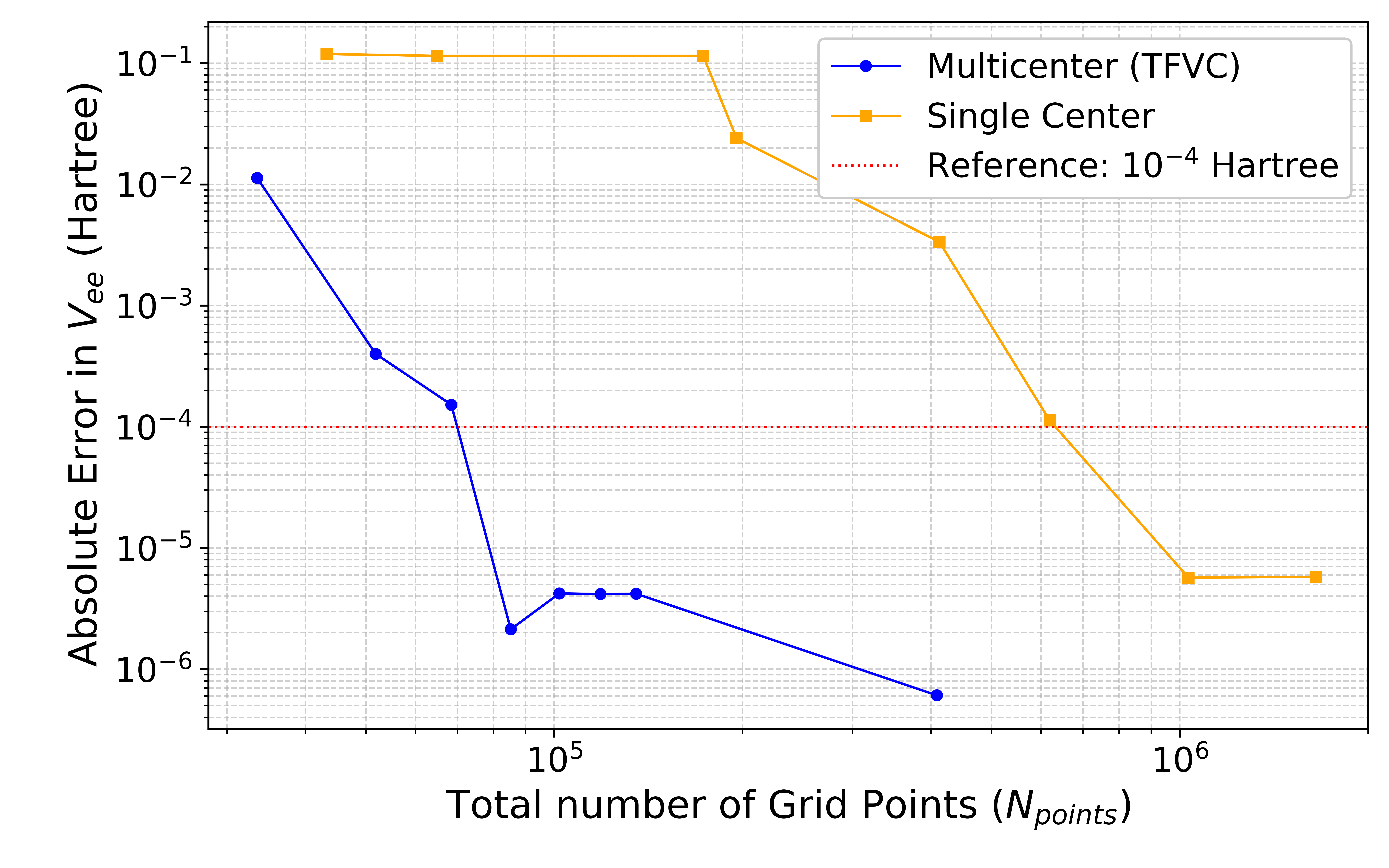}
\caption{Convergence of the electron-electron repulsion energy ($V_{ee}$) for propane ($\text{C}_3\text{H}_8$) as a function of the total number of grid points. The plot compares the performance of the standard single-center expansion against the current multicenter integration scheme. See table S1 for the details on the grid used. }
\label{fig:propane_convergence}
\end{figure}

As shown in Figure~\ref{fig:propane_convergence}, the single-center method exhibits significant convergence difficulties because it requires extremely high angular resolution to capture density features located far from the coordinate origin.
Even when employing the largest available Lebedev grids, with 5810 angular nodes and more than $1 \cdot 10^6$ effective integration points after symmetry reduction (I(\textbf{u})=I(\textbf{-u})), the single-center integration error remains an order of magnitude higher than that achieved by the multicenter scheme. In contrast, the multicenter method rapidly reaches high accuracy, with a relative error below $0.01\%$, using a modest grid of only 590 angular nodes per center.\newline

\begin{figure}[h]
    \centering
    \includegraphics[width=1\linewidth]{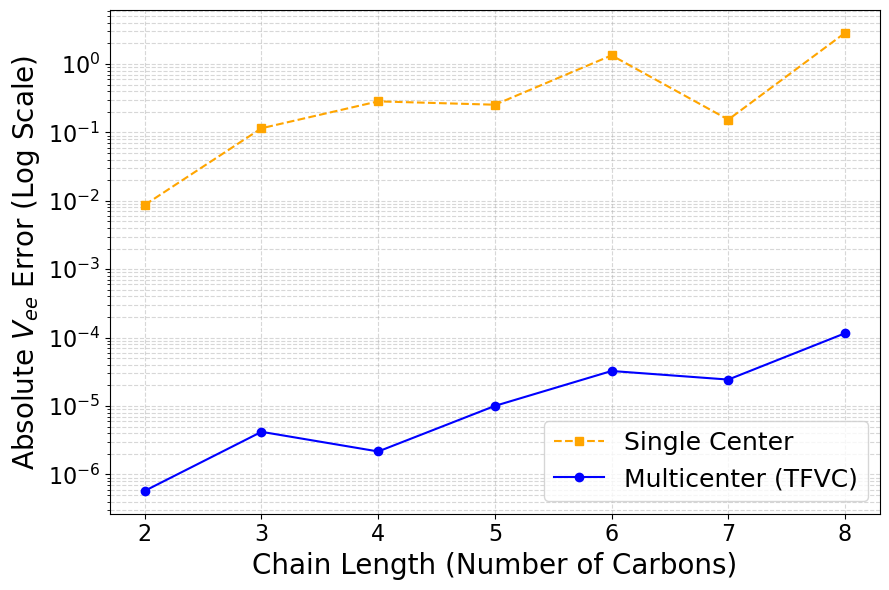}
    \caption{Absolute error in the calculated $V_{ee}$ for the linear alkane series, from ethane to octane. The total grid size was constrained to approximately 300,000 points for the single-center method and 250,000 points for the multicenter method. See Table S2 for the specific grids used in these calculations.}
    \label{fig:alkanes}
\end{figure}

To compare the performance of the single-center and multicenter integration schemes as molecular size increases, we calculated $V_{ee}$ for the linear alkane series, from ethane to octane. For the single-center approach, we utilized a fixed grid of 1016 radial and 590 angular nodes, yielding approximately 300,000 total points for all the molecules (after accounting for intracule symmetry). For the multicenter method, the angular grid was kept constant at 590 nodes per center, while the number of radial nodes per center was systematically decreased (from 338 for ethane down to 48 for octane). This purposely capped the multicenter grid at approximately 250,000 total points across the entire series. As shown in Figure~\ref{fig:alkanes}, while the integration error naturally increases with system size for both methodologies, the multicenter scheme is systematically four orders of magnitude smaller than the single-center one. Despite using roughly 50,000 fewer total grid points, the multicenter scheme keeps the absolute error below $10^{-4}$ for the largest system. In contrast, as the molecular chain lengthens, the origin-anchored single-center grid becomes effectively blind to the increasingly distant electron-pair peaks, resulting in extremely large numerical error. 

Having established the superiority of the multicenter integration scheme over the single-center approach, we next assess its performance on a larger statistical sample using a subset of the S22 dataset,\cite{juerevcka:06pccp} as described in Table S3. This set covers a broad range of structural motifs, from simple hydrides and hydrogen-bonded clusters to large dispersion-dominated systems. All baseline calculations were performed using 50 radial and 590 angular nodes per integration center.

\begin{figure}[h]
    \centering
    \includegraphics[width=0.92\linewidth]{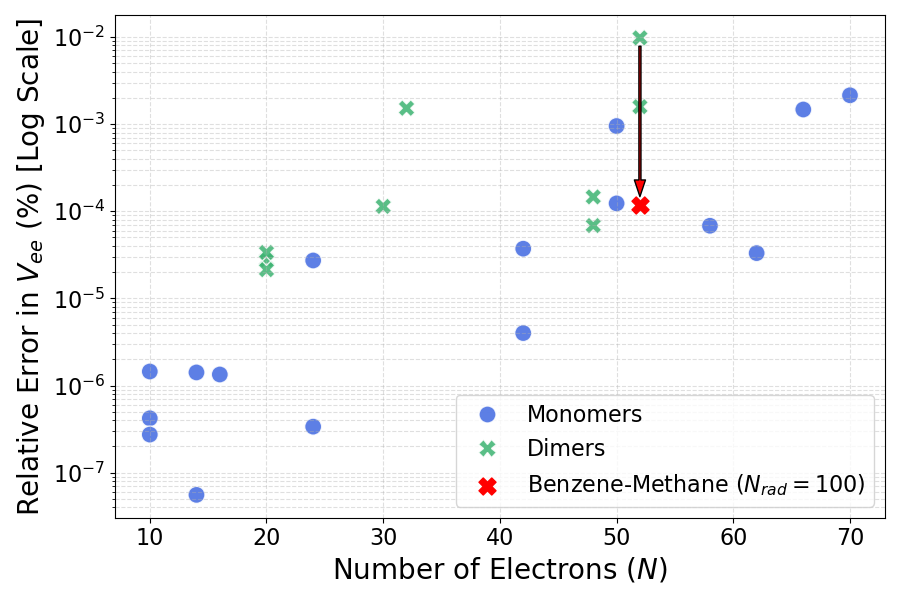}
    \caption{Relative integration error percentage in $V_{ee}$ as a function of the number of electrons for the S22 dataset, using 50 radial and 590 angular nodes per integration center. Results are shown for the constituent monomers (blue circles) and their corresponding dimer complexes (green crosses) using the TFVC multicenter integration scheme.}
    \label{fig:s22}
\end{figure}

Figure \ref{fig:s22} illustrates the relative integration error in $V_{ee}$ as a function of the total number of electrons. Although $V_{ee}$ values span hundreds of Hartrees, the relative error of the TFVC method remains exceedingly small, with the majority of molecules having errors below 
$0.0001\%$. 

The absolute error grows predictably with system size, remaining well within acceptable tolerances. For the monomer set, the mean absolute error (MAE) in $V_{ee}$ is a mere 1.74 mHartree, while for the dimer complexes, the MAE is 4.97 mHartree. The largest absolute deviation observed using the baseline grid is approximately 35.4 mHartree, which corresponds to the challenging benzene-methane dispersion complex ($N = 52$ electrons). Even for this outlier, the relative error remains below $ 0.01\%$ and, as also shown in Figure~\ref{fig:s22}, increasing only the radial grid density from 50 to 100 nodes rapidly drops the relative error by nearly two orders of magnitude.

This result highlights a fundamental advantage of the multicenter TFVC approach over single-center methods: angular convergence is largely decoupled from the spatial extent of the system. In a single-center framework, capturing electron-pair density far from the origin requires a very large number of angular grid points to avoid undersampling the long-range regions of the intracule. In contrast, the multicenter approach reliably achieves angular convergence with a universal baseline of 590 nodes per center, independently of the overall molecular size or topology. When tighter accuracy is required for challenging long-range tails, the remaining error can be systematically reduced by simply extending the radial grid. A comparable improvement in a single-center scheme would generally require increasing both the radial and angular resolutions, leading to a much larger computational cost.


\subsection{Visualization of radial intracule using Kernel Density Estimation}

Although the primary focus of this work is to achieve efficient integration over the intracular coordinate for the evaluation of scalar properties, in this section we discuss the radial intracule distribution, $I(u)$, because of its importance in characterizing the spatial extent of electron correlation. This quantity can, in principle, be obtained using standard single-center expansions followed by integration over the angular coordinates. However, such an approach suffers from an \textit{angular sparsity} problem: as the interelectronic distance $u$ increases, the surface density of a fixed angular quadrature decreases as $1/u^2$, leading to numerical noise and integration errors at long range.

To obtain $I(u)$, we apply the proposed KDE technique to two particularly challenging systems that suffer from the aforementioned angular sparsity problem: the helium dimer ($\mathrm{He}_2$) at an internuclear separation of $R=20$ bohr and the formic acid dimer.

\begin{figure}[H]
    \centering
\includegraphics[width=0.95\linewidth]{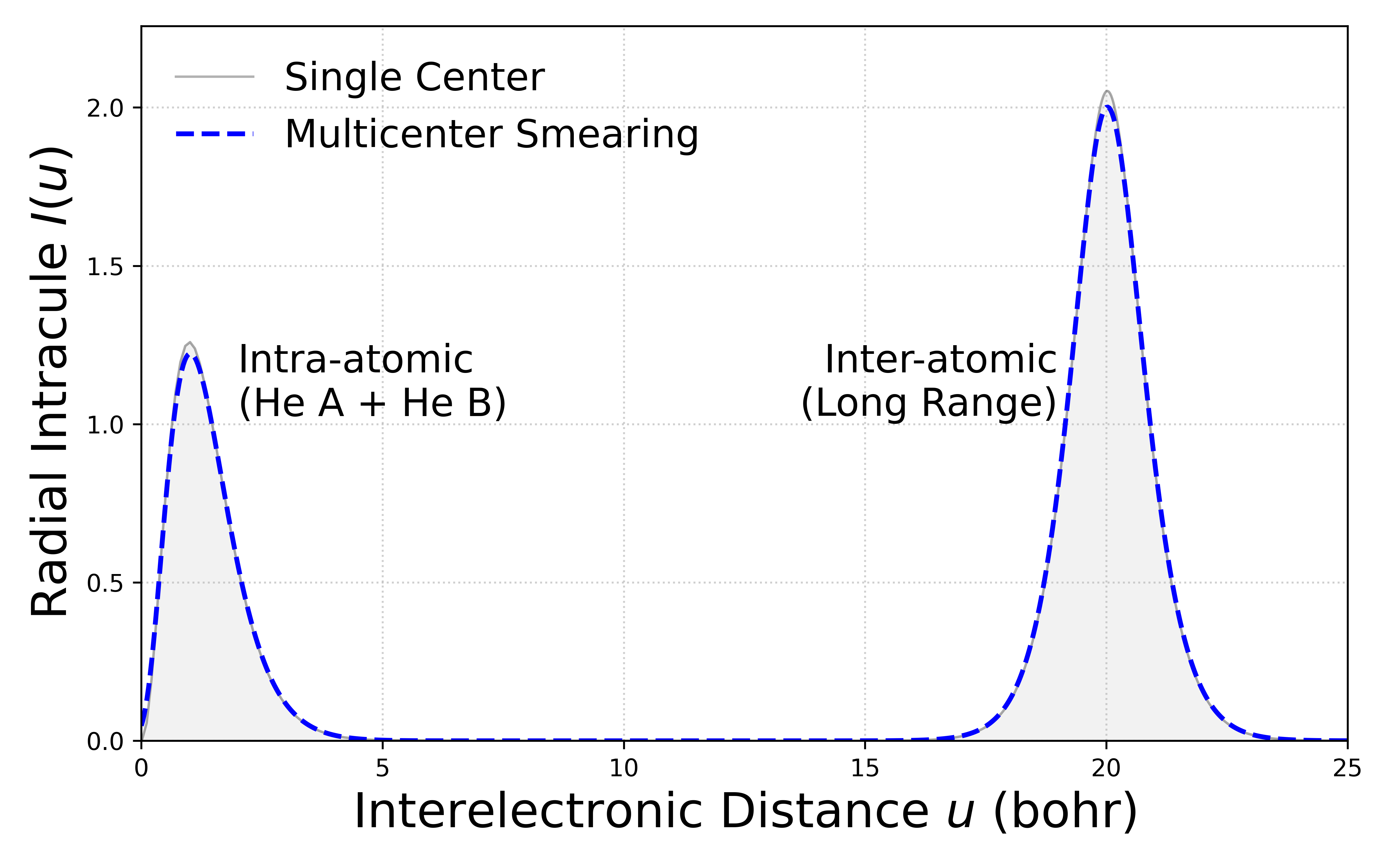}
    \caption{Radial intracule $I(u)$ for the dissociated helium dimer ($R_{AB}=20$ bohr), comparing the reference single-center integration (grey area) and the proposed multicenter Gaussian smearing method (blue dashed line).}
    \label{fig:he2_reconstruction}
\end{figure}

$\mathrm{He}_2$ exhibits both short-range intra-atomic electron pair peak at $u \sim 1$ bohr and an extreme long-range interatomic pair peak at $u \sim 20$ bohr (Figure~\ref{fig:he2_reconstruction}). To capture this intracule profile, the standard single-center integration requires a massive, highly dense numerical grid (1057 radial $\times$ 5810 angular nodes, totalling $\approx 3.07 \times 10^6$ points) to yield 6.0011 electron pairs. In contrast, the approximated multicenter (TFVC) integration drastically reduces the computational cost by partitioning the space into local grids. Using just 3 centers (the origin and the two interatomic vectors) with 200 radial and 590 angular nodes per center ($\approx 1.8 \times 10^5$ total points), the KDE method successfully recovers both the short-range ($u \sim 1.0$ bohr) and long-range ($u \sim 20.0$ bohr) peaks. This represents a nearly 17-fold reduction in total grid size while retaining a highly accurate integrated pair count of 5.9958. Although the KDE smearing slightly attenuates the absolute maximum of the sharpest peaks, it perfectly preserves the physically meaningful topological features of the distribution.

\begin{figure}[h]
    \centering
    \includegraphics[width=0.95\linewidth]{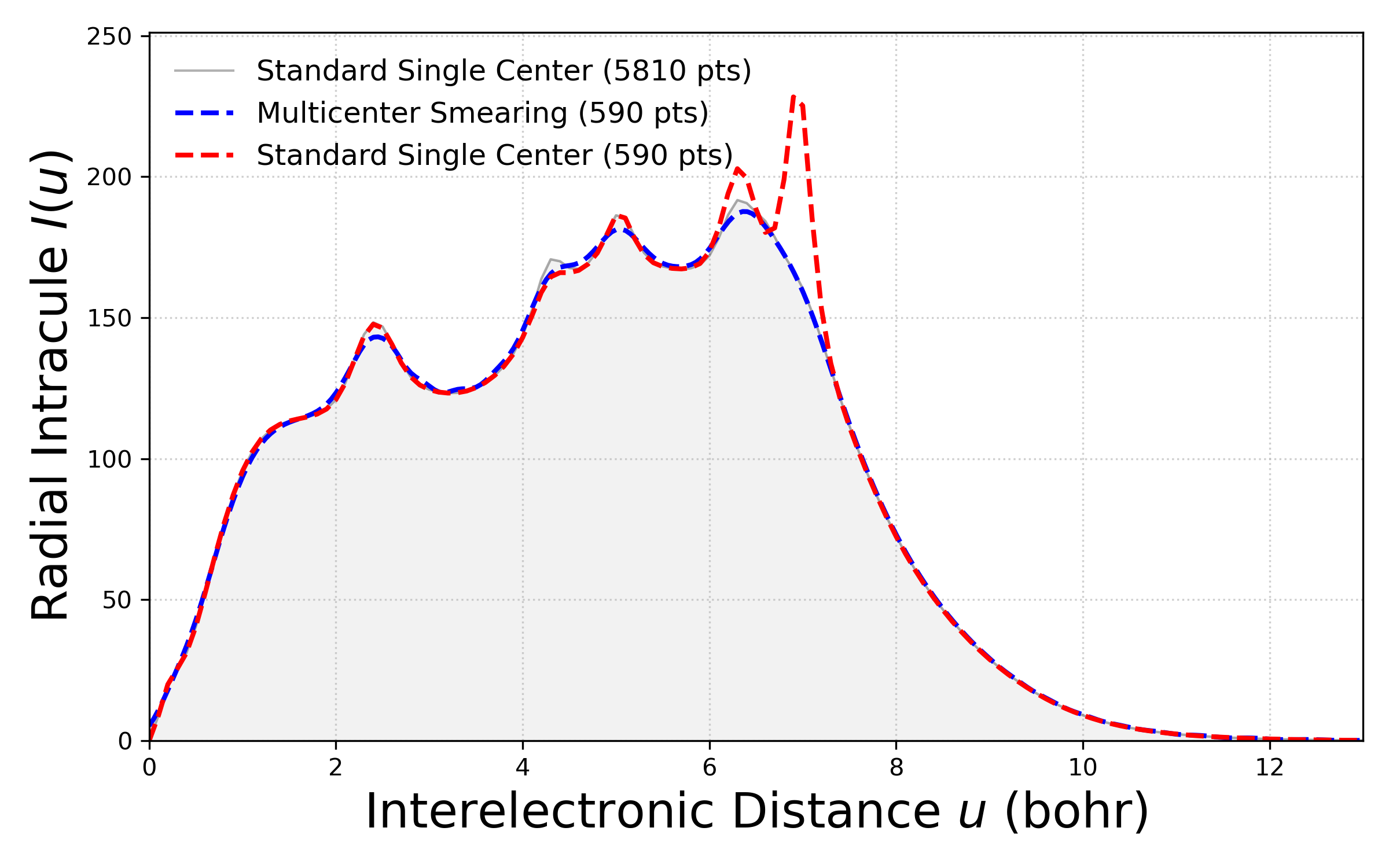}
    \caption{Radial intracule for the formic acid dimer. The plot compares the reference distribution (grey area), a standard 590-point single-center integration (red dashed line), and the multicenter Gaussian KDE reconstruction (blue dashed line).}
\label{fig:formic_dimer_reconstruction}
\end{figure}

The formic acid dimer presents a much more complex topology due to the multitude of distinct interacting interatomic pairs (Figure~\ref{fig:formic_dimer_reconstruction}).

A single-center expansion using a standard angular grid (133 radial $\times$ 590 angular nodes, totaling 39,235 points) suffers from severe numerical artifacts, generating a spurious, unphysical peak around $u \sim 7$ bohr (dashed red line). This occurs because the fixed number of angular points becomes vastly too sparse to sample the expanding spherical volume accurately. To correctly reproduce these long-range features within a single-center approach, an impractically dense grid is required (133 radial $\times$ 5810 angular nodes, totaling 386,365 points). Such a steep scaling makes evaluating long-distance radial nodes computationally prohibitive.

In contrast, the multicenter KDE volume-projection approach (dashed blue line) completely avoids these artificial long-range oscillations by using a highly scalable grid (19 centers $\times$ 60 radial $\times$ 590 angular nodes, totaling 336,300 points). Despite using 50,000 fewer total points than the converged single-center reference, it correctly recovers the qualitative topological structure of the intracule. The distribution accurately positions the local maxima corresponding to the intra-atomic core ($u < 2$ bohr), the covalent network ($2 < u < 5$ bohr), and the long-range intermolecular interactions ($u > 5$ bohr).

The mathematical robustness of the KDE method is definitively confirmed by integrating the radial probability distribution. For the formic acid dimer ($N=48$), the exact analytical number of electron pairs is $N(N-1)/2 = 1128$. The unphysical 590-node single-center calculation deviates significantly, yielding 1148.15 pairs due to the long-range artifacts. The expensive 5810-node single-center calculation yields 1127.76 pairs. Remarkably, the multicenter KDE approach yields 1127.46 pairs, proving that it reliably conserves the total electron pair probability while entirely bypassing the catastrophic angular scaling of standard expansions.

\subsection{Range-Separated Analysis of Electron Correlation and London Dispersion interactions}

Since our multicenter TFVC scheme partitions the total integral over interatomic vectors, $\mathbf{u}_{AB}$, the electron–electron repulsion energy, $V_{ee}$, as well as other two-electron properties, can be decomposed into short- and long-range components. This capability provides a unique tool for analyzing electron correlation. In particular, our group has previously identified the long-range part of the so-called $c_{II}$ component of the Coulomb hole ---defined as the difference between the intracule of a correlated wavefunction and its Hartree–Fock counterpart\cite{via-nadal:19jpcl}--- with the dispersion energy.\cite{via-nadal:17pra,via-nadal:19jpcl,via-nadal:22jcp} Accordingly, integrating the intracule center associated with a pair of atoms bound by dispersion interactions yields a quantity proportional to the dispersion energy.

\begin{figure}[h]
\centering
\includegraphics[width=0.95\linewidth]{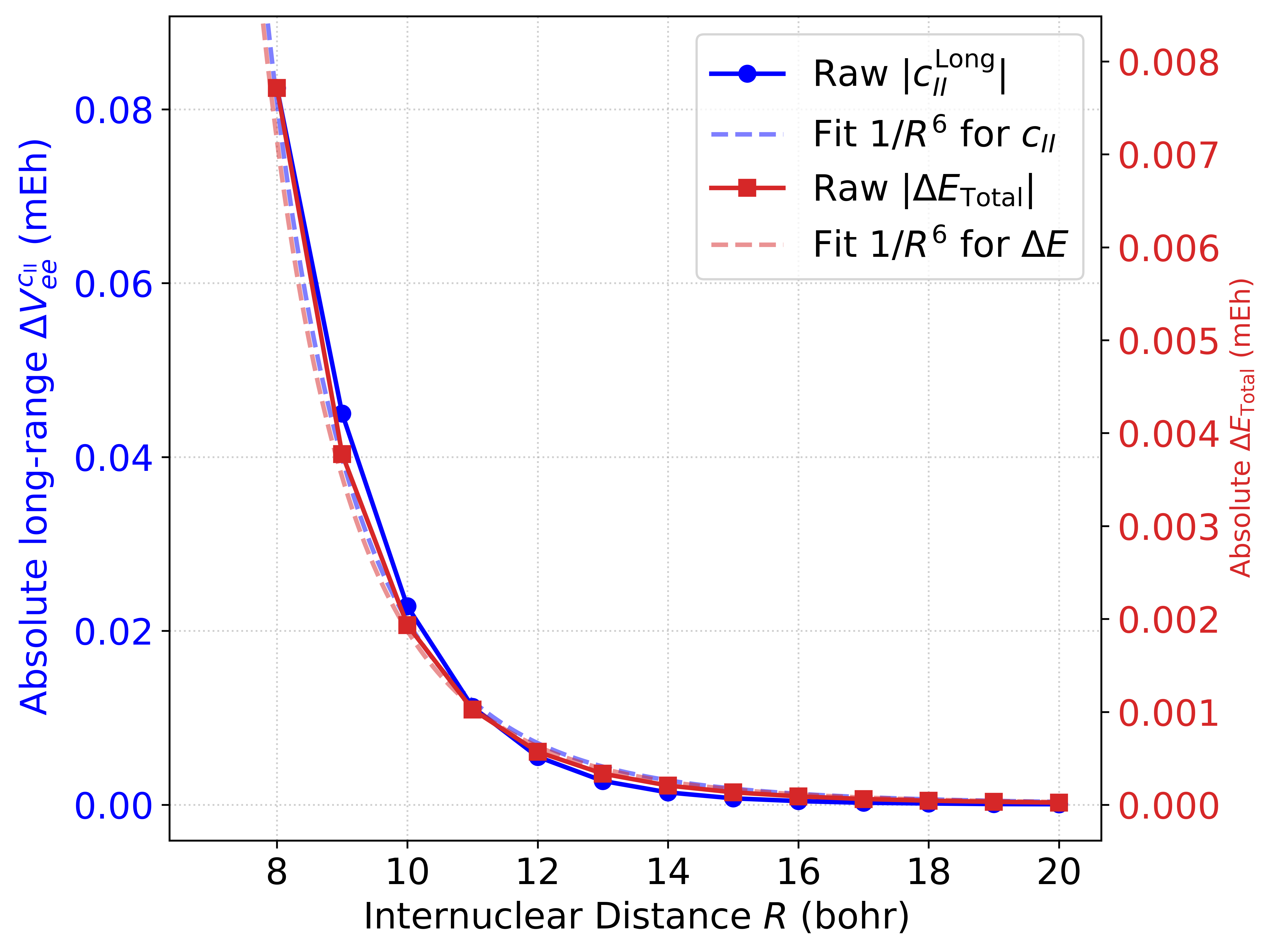} 
\caption{Comparison of the long-range asymptotic decay of the long-range $\Delta V_{ee}^{\text{c}_\text{II}}$ (blue circles) and the dissociation energy ($|\Delta E_{\text{dis}}|$, red squares) for the He$_2$ dimer. Dashed lines represent theoretical $1/R^6$ decay curves. Data are plotted on two different linear scales to show the similar behaviour regardless the magnitude.}
\label{fig:he2_dispersion_decay}
\end{figure}

\begin{figure}[h]
\centering
\includegraphics[width=0.95\linewidth]{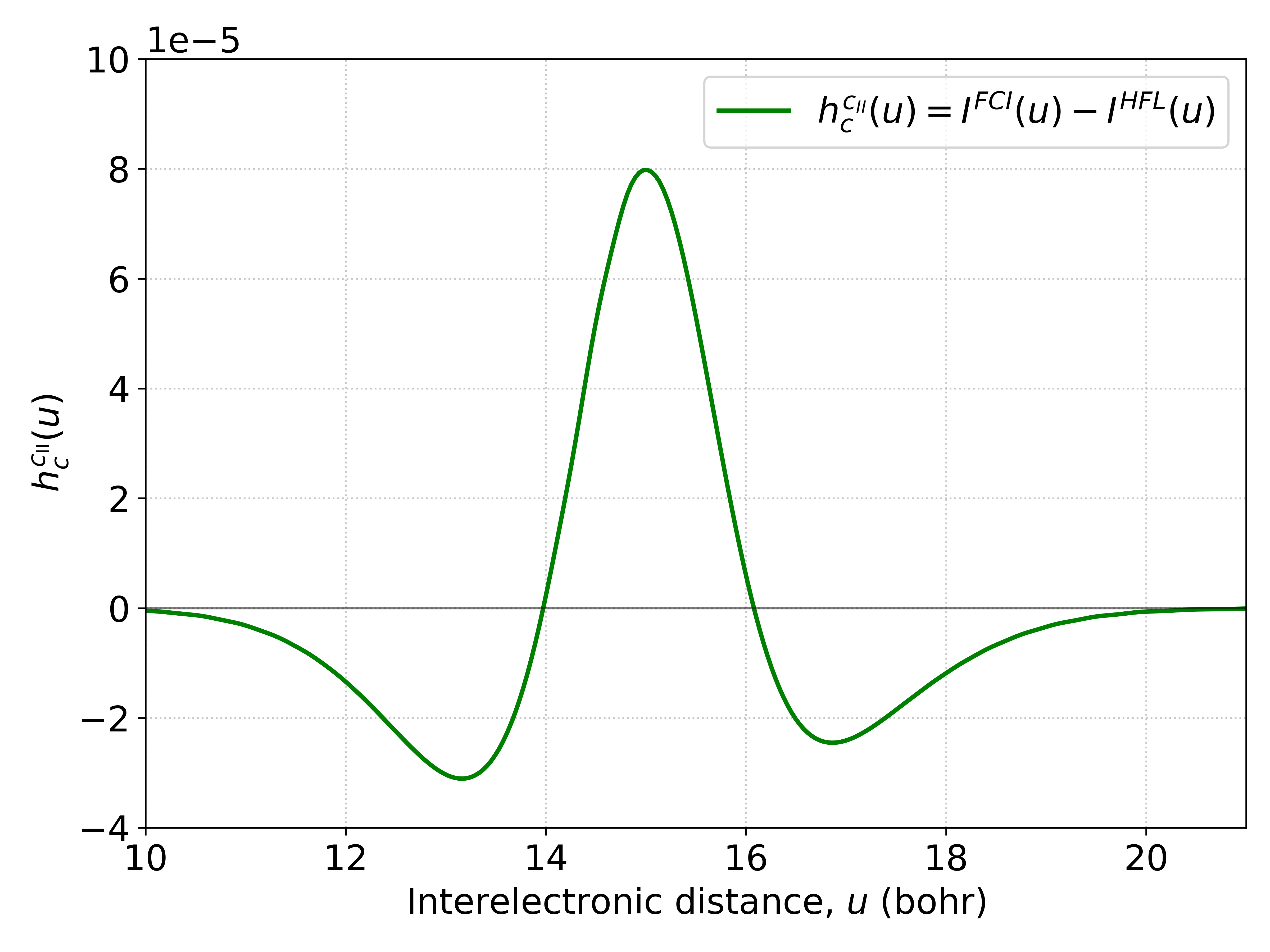} 
\caption{Profile of the long-range inter-atomic $c_{II}$ Coulomb hole component for the He$_2$ dimer separated by 15.0 bohr, using the KDE smearing method from a multicenter grid. The curve represents the difference in the radial intracule probability distribution ($\Delta I(u)$) obtained from full configuration interaction (FCI) and Hartree-Fock-like (HFL) calculations, using the aug-cc-pVTZ basis set. The x-axis is restricted to the inter-atomic region ($u \in [10.0, 20.0]$ bohr) to isolate the long-range correlation features that correspond to London dispersion interactions.}
\label{fig:KDE_dispersion}
\end{figure}

To assess the capabilities of the present multicenter algorithm, we selected a particularly challenging system: the helium dimer, whose dispersion interaction is extremely weak. If the method is able to recover the characteristic $1/R^6$ energy decay, this would demonstrate its accuracy in retrieving even the smallest London dispersion interactions. As shown in Figure~\ref{fig:he2_dispersion_decay}, the present algorithm indeed recovers the expected energy decay, which we compare with the dimer binding energy. Details of the procedure used to obtain the $c_{II}$ component are provided in the Supporting Information. Finally, Figure~\ref{fig:KDE_dispersion} also shows that the KDE method reproduces the main features of the long-range $c_{II}$ component of the Coulomb hole, demonstrating the robustness of the method even for small-magnitude interactions that are difficult to capture with one-center integration schemes.

\section{Conclusions}

In this work, we have presented and validated an adapted multicenter numerical integration scheme for the calculation of radial intracule densities and their associated moments. By combining Becke’s geometric partitioning with Topological Fuzzy Voronoi Cell (TFVC) weight functions, we have developed a robust, parameter-free approach that overcomes the limitations of standard single-center expansions. Whereas single-center grids suffer from \textit{blind spots} in the bonding regions, the multicenter distribution naturally clusters quadrature points around the secondary maxima of the intracule density, as confirmed by the topological analysis of the intracule function. Numerical evidence is provided for the alkaline-earth series and the S22 dataset.

The topological partitioning of intracule space can now be fully exploited by decomposing two-electron properties according to interelectronic vectors. For instance, $V_{ee}$ can be partitioned into short-, medium-, and long-range contributions, and the integration of the latter can be employed to characterize dispersion interactions.

We have also introduced a volume-projection technique based on Gaussian Kernel Density Estimation (KDE) to reconstruct the radial intracule distribution, $I(u)$, from a multicenter grid. This approach effectively circumvents the severe angular sparsity problem—the $1/u^2$ decay of the surface density—that affects standard surface integration at large interelectronic distances. By treating the multicenter grid as a discrete volume sampling and applying Gaussian smearing, smooth and qualitatively accurate distributions can be recovered even for extremely long-range correlation peaks.

In summary, this adapted multicenter strategy effectively removes the integration bottleneck that has historically hindered the application of intracule analysis to extended systems and has the advantage of being applicable to arbitrary basis sets. The latter is particularly relevant for the development of new electronic structure methods tailored for electron correlation.\cite{paulau:25jctc}\newline 

\section{Acknowledgements}
We are grateful for financial support from the Ministerio de Ciencia e Innovación (PID2025-170025NB-C31) provided by MCIN/AEI/10.13039/501100011033 and FSE+ and the Basque Government (project IT-2067-26). The authors acknowledge the computational resources and technical and human support provided by the DIPC. M.Y. acknowledges PIF2021 grant from Euskal Herriko Unibertsitatea. 

\bibliography{biblio,gen}
\end{document}